# UNIDIRECTIONAL MAGNETOSTATIC WAVES

## Lock E.H.* and Vashkovsky A.V.

Institute of the Radio Engineering and Electronics RAS, Fryazino' department.

Dispersion characteristics of magnetostatic waves in tangentially magnetized to saturation ferrite film with a "magnetic wall" condition (tangential component of microwave magnetic field is equal to zero) on the one of the film surface were calculated. It is found, that unidirectional magnetostatic waves are appeared in this geometry: they can transfer energy only in one direction and fundamentally cannot transfer energy in an opposite direction.

* The author is also known by the name Lokk E. G. which was appeared as a result of transliteration under the BSI/ANSI scheme adopted by many journals.

Magnetic materials, in particular ferrites, are characterized by anisotropic properties and various energetic interaction, such as dipole, exchange, magnetoelastic and magneto-optical, so electromagnetic waves in ferrites pretend to be in the lead as an object, that is used for investigation and realization of many unconventional physical effects and phenomena. For example, a negative refraction can occur in ferrite film structures not only for the earlier predicted case where the incident wave is forward and refracted wave is backward, but also in the case where both waves are forward [1 - 3]. Moreover, as a distinction from isotropic media there is possible situation in these structures where a wave incident normally to the boundary deflects from the normal to the boundary after refraction [1 - 3]. Since a technology of creation of composite materials is developed at last years, so it is reasonable that new fundamental effects may be appear in new artificial media, created by using of ferrites together with composite materials. Properties of electromagnetic waves in these media may be extraordinary in principle. In particular, the using of composite materials, that can provide the "magnetic wall" condition (i.e. tangential microwave magnetic field components are equal to zero) on the one of the ferrite film surface, result in appearance of "unidirectional" dipole spin wave in such structure. We name this wave "unidirectional" because it can transfer energy only in certain plane direction and the wave transferring energy in opposite direction does not exist in the structure. Dispersion dependence and properties of this wave are described below.

Consider eigenmode propagation in an infinite thin ferrite film (slab) of thickness $d$. Let film is magnetized to saturation by uniform magnetic field $\overrightarrow{H_0}$, directed along the $z$ axis ($z$ and $y$ axes lie in the film plane and the $x$ axis is perpendicular to the one (Fig.1)). The magnetic permeability of ferrite is characterized by tensor:

$$\overleftrightarrow{\mu} = \begin{vmatrix} \mu & -i\nu & 0 \\ i\nu & \mu & 0 \\ 0 & 0 & 1 \end{vmatrix}, \tag{1}$$

where

$$\mu = 1 - \frac{\omega_M \omega_H}{\omega^2 - \omega_H^2}, \tag{2}$$

$$\nu = \frac{\omega_M \omega_H}{\omega - \omega_H^2}, \tag{3}$$

$\omega_H = \gamma H_0$, $\omega_M = 4\pi\gamma M_0$, $\omega = 2\pi f$, $\gamma$ is the gyromagnetic constant, $4\pi M_0$ is the ferrite saturation magnetization and $f$ is the electromagnetic oscillation frequency. It is assumed that magnetic permeability of around half-spaces is equal to unity.

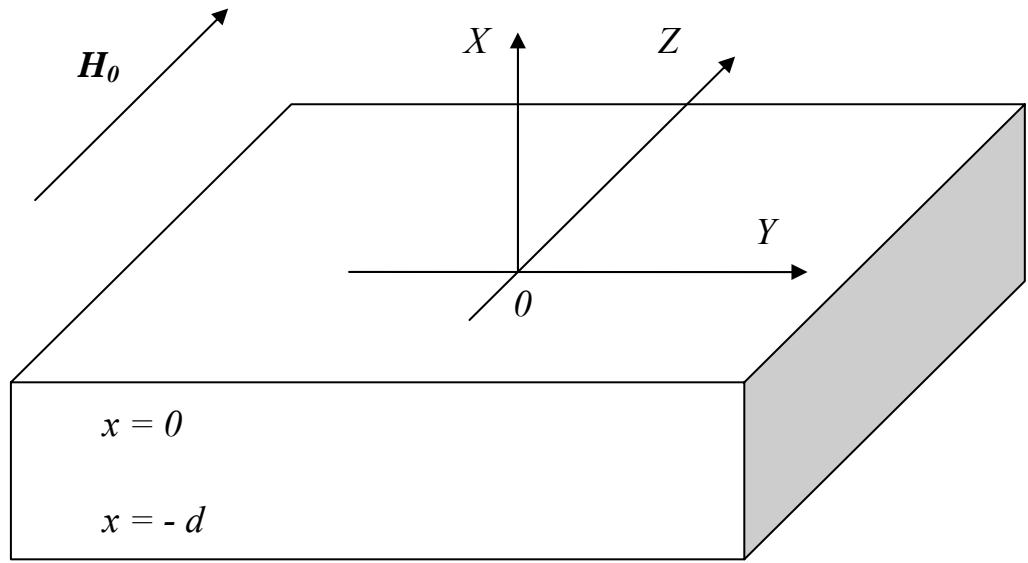

Fig. 1. Geometry of the problem.

Dipole spin waves or so called magnetostatic waves (MSW) can propagate with small losses in tangentially magnetized ferrite film. The phase velocity of these waves is much fewer than the light velocity $c$ and is much larger then the velocity of exchange spin waves. Due to this property of MSW we can neglect by both the terms with $\partial/\partial t$ in Maxwell's equations and the terms of exchange energy influence. Thus to calculate characteristics of MSW we can use equations $rot(\vec{h}) = 0$ and $div(\vec{b}) = 0$, where $\vec{h}$ and $\vec{b}$ are microwave magnetic field and induction respectively, and introduce magnetostatic potential $\psi$:

$$\vec{h} = \nabla \psi \qquad\qquad (4)$$

As it is known two types of MSW can be excited in the free ferrite film: backward volume MSW propagating within certain sector of angles near $z$ axis, and surface MSW propagating within another sector of angles near $y$ axis (Fig.1) [4]. Dispersion characteristics of MSW were also studied in detail for the film, bounding with conducting plane [5, 6] and the film, surrounded by half-spaces with negative permittivity [7]. The case of the film, bounding with "magnetic wall", was not investigated yet.

"Magnetic wall" was a theoretic abstraction some time ago and did not represent an object of practical interest. However different properties of composite materials were studied little by little and the material imitating "magnetic wall" and providing the $h_t = 0$ as a boundary condition on its surface is appear to be possible (see, for example [8]). We look forward that the materials with the same property will design to provide "magnetic wall" condition on the ferrite film surface.

We'll describe below MSW dispersion characteristics in ferrite film with "magnetic wall" condition on the one of the film surface and will show, that paradox situation is realized for this geometry: MSW becomes unidirectional wave at all frequency range, where it exists (As it is known MSW demonstrates unidirectional property also in metallized ferrite films at the part of the frequency range $\omega_H + \omega_M/2 < \omega < \omega_H + \omega_M$ ).

Let's consider properties of MSW in ferrite film under "magnetic wall" condition for the very simple case, when the wave vector $\vec{k}$ is parallel to $y$ axis. As it is seen from (4) microwave magnetic field $\vec{h}$ is determined by the magnetic potential $\psi$. Within the film the potential is described by expression

$$\psi_i = (A \exp(-kx) + B \exp(kx)) \exp(\pm i\, ky) \exp(i\omega t), \qquad (5)$$

and outside of the film – by expression

$$\psi_e = C \exp(kx) \exp(\pm i\, ky) \exp(i\omega t), \qquad (6)$$

where $k = \left| \vec{k} \right|$ and $A, B, C$ are coefficients.

Suppose the upper surface of the film ($x = 0$) is under "magnetic wall" condition. i.e. $h_{iy} = 0$ (for $x = 0$). At the lower surface, bonding with free space, another boundary conditions take place: $\psi_i = \psi_e$ , $b_{ix} = b_{ex}$ (for $x = -d$). Following by the method used in [4], composing and solving the system of equations one can find the next dispersion relationship:

$$\text{th}(kd) = \frac{-\mu}{1 \pm \nu} \qquad (7)$$

Sign "+" in (7) corresponds to the case, when the $\vec{k}$ vector is directed along $y$ axis ($k$ is positive), and sign "-" in (7) – to the case, when the $\vec{k}$ vector is directed oppositely regarding to $y$ axis ($k$ is negative). In the following analysis we use definitions of the phase $\overrightarrow{v_{ph}}$ and group $\overrightarrow{v_g}$ velocities in accordance with formulas $\overrightarrow{v_{ph}} = \omega / \vec{k}$ и $\overrightarrow{v_g} = \partial \omega / \partial \vec{k}$ respectively. Dispersion dependences $f(k)$ calculated from (7) are shown in Fig. 2, where boundary frequencies are also indicated. One can see in Fig. 2 that the upper curve is not distinct essentially from the one for the free ferrite film: it is positioned at the same frequency interval and is described forward surface MSW ($\overrightarrow{v_{ph}}$ and $\overrightarrow{v_g}$ are codirectional vectors). But in Fig. 2 there is not second upper curve, that exists for the geometry of free ferrite film and is symmetric to the first curve regarding to the frequency axis! Thus every frequency corresponds with only one value of the wave number (negative value) and the group velocity $\overrightarrow{v_g}$ of the forward surface MSW is always directed oppositely to $y$ axis and can not be directed along $y$ axis in principle, i.e. this MSW is unidirectional.

*f*, GHz

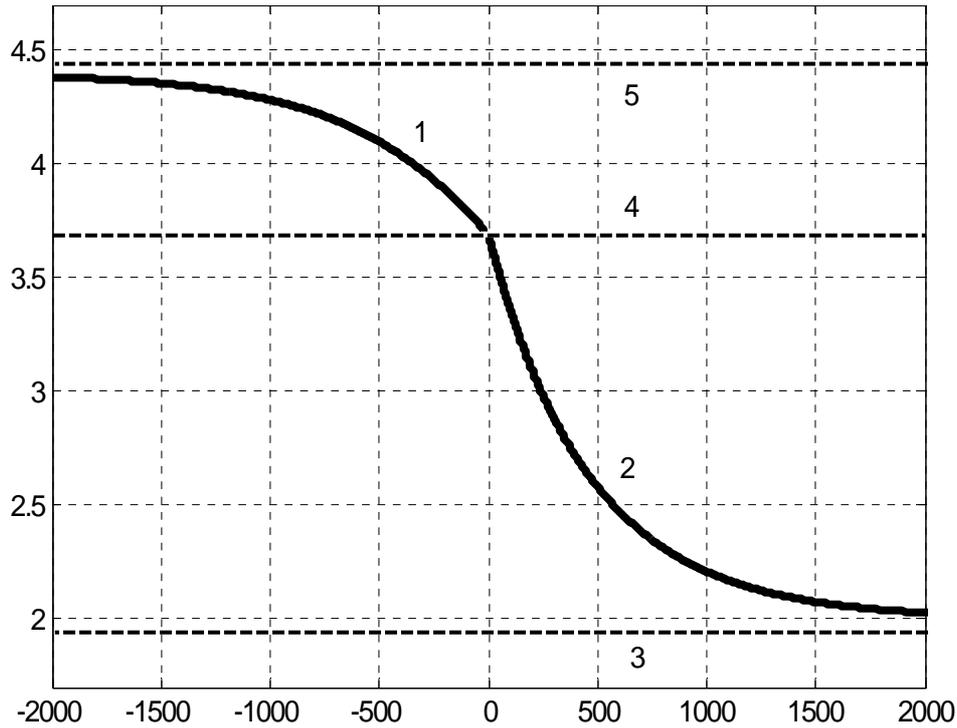

*k*, cm$^{-1}$

Fig. 2. Calculated dependences *f(k)* obtained at $H_0$ = 700 Oe, $d$ = 10 μm, $4\pi M_0$ = 1750 Gs: 1 – forward surface MSW, 2 – backward surface MSW. Straight dashed lines 3 – 5 show boundary frequencies $\omega_H/2\pi$, $\sqrt{\omega_H^2 + \omega_H \omega_M}/2\pi$ and $(\omega_H + \omega_M/2)/2\pi$ respectively.

What is surprisingly indeed, that is appearance of the lower curve in the Fig.2. This curve corresponds to the surface MSW too, but it is positioned at the frequency interval $\omega_H < \omega < \sqrt{\omega_H^2 + \omega_H \omega_M}$ where volume MSW (described by potential $\psi_i \sim A \sin(kx) + B \cos(kx)$) does exist in the case of free ferrite film.

Moreover this surface MSW is backward. In Fig. 2 there is not second lower curve that would be located symmetrically regarding to the frequency axis and so every frequency corresponds with only one value of the wave number (positive value). Thus this wave is unidirectional too, its group velocity is always directed oppositely to $y$ axis and can not be directed along $y$ axis in principle. Mention must be made that in the geometry of free ferrite film the backward wave propagating parallel to $y$ axis does not exist [4].

The direction of energy transport (direction of group velocity $\overrightarrow{v_g}$) for both types of MSW is determined by the direction of applied uniform magnetic field $\overrightarrow{H_0}$. If, for instance, $\overrightarrow{H_0}$ vector indicated in the Fig.1 will change its orientation to opposite then MSW will transport energy only in positive direction of $y$ axis.

Thus, backward surface MSW in frequency interval $\omega_H < \omega < \sqrt{\omega_H^2 + \omega_H \omega_M}$ and the forward surface MSW in frequency interval $\sqrt{\omega_H^2 + \omega_H \omega_M} < \omega < \omega_H + \omega_M/2$ can be propagated in ferrite film (slab) having the boundary condition of "magnetic wall" on the one of its surfaces. Transport of energy by both waves is unidirectional.

Investigation of unidirectional waves opens up not only new possibilities for design of analog microwave devices, but defines and puts forward new problems, such as study of refraction and diffraction of these waves.

This work was supported by the Russian Foundation for Basic Research, project no. 04-02-16460, and the Program of Fundamental Research of the Russian Academy of Sciences "Investigation of Electrophysical Effects Observed during Electromagnetic Energy Flux Propagation in Metamaterials".

REFERENCES

1. Vashkovskii A.V., Lokk E.G. *Usp. Fiz. Nauk* **174** (6) 657 (2004) [Physics – Uspekhi **47** (6) 601(2004)].
2. Vashkovskii A.V. et al. *Radiotekh. Elektron.* **36** 1959 (1991)
3. Vashkovskii A.V. et al. *Radiotekh. Elektron.* **36** 2345 (1991)
4. Damon R. W., Eshbach J. R., *J. Phys. Chem. Solids* **19** (3/4) 308 (1961).

5. Van de Vaart H. *Electronics Lett* **6** 601 (1970).
6. Bongianni W. L. *J. Appl. Phys*. **43** 2541 (1972).
7. Vashkovskii A. V., Lokk E. G., *Radiotekh. Elektron.* **47**, 97 (2002) [J. Commun. Technol. Electron. **47**, 87 (2002)].
8. Sievenpiper D.F. Dissertation for the degree of Doctor of Philosophy in Electrical Engineering. Los Angeles, University of California, 1999.